\documentclass[12pt]{article}

\usepackage{amsmath}
\input epsf

\numberwithin{equation}{section}

\def\bfone{\relax{\rm 1\kern-.36em 1}}
\def\zero{\relax{ 0\kern-.38em 0}}
\def\ZZ{\relax{Z\kern-.50em Z}}

\newcommand{\CF}{ {\cal F} }
\newcommand{\CG}{ {\cal G} }
\newcommand{\CN}{ {\cal N} }
\newcommand{\CZ}{ {\cal Z} }

\newcommand{\hv}{ h^{\vee}  }
\newcommand{\av}{ \alpha^{\vee}  }
\newcommand{\AD}{ A_\mathrm{D}  }
\newcommand{\gc}{ \mathsf{g}_{\,\mathrm{C}}  }

\begin{document}

\addtolength{\baselineskip}{2pt}
\thispagestyle{empty}

\vspace{2.5cm}

\begin{center}
{\scshape\Large 
Charges of dyons in $\CN=2$ supersymmetric \\ gauge theory}

\vspace{1.5cm}

{\scshape\large Michael Yu. Kuchiev}

\vspace{0.5cm}
{\sl School of Physics, University of New South Wales,\\
Sydney, Australia}\\
{\tt kmy@phys.unsw.edu.au}\\
\vspace{1cm}

{\Large ABSTRACT}

\vspace{0.3cm}

\end{center}

Expressions for electric and magnetic charges of dyons, which become massless in the strong-coupling limit of the supersymmetric $\CN=2$ gauge theory with an arbitrary gauge group are presented. Transitions into different vacua of the $\CN =1$ gauge theory, when the $\CN=2$ supersymmetry is broken explicitly to the  $\CN =1$ case, are discussed. 
The existence of a minimal set of light dyons, which are necessary to describe this transition, is established. The total number of these dyons equals the product of the rank and dual Coxeter number of the gauge group. A conjecture, which states that this minimal set incorporates all possible light dyons, is discussed. A relation of dyon charges with monodromies at weak and strong couplings is outlined and comparison with known charges of dyons for particular gauge groups is made.

\newpage

\section{Introduction}
\label{intro}

The properties of dyons, which become massless in the strong-coupling limit
in the pure $\CN=2$ gauge theory described by the Seiberg-Witten solution are discussed. Explicit simple expressions for magnetic and electric charges of these dyons are written for an arbitrary gauge group. The total number of dyons is shown to depend on two parameters that govern the gauge algebra, its dual Coxeter number and its rank. The number of different massless dyons is shown to be related to the Witten index, which equals the number of different vacua in the $\CN =1$ supersymmetric gauge theory.

The Seiberg-Witten solution for the $\CN=2$ supersymmetric gauge theory \cite{Seiberg:1994rs,Seiberg:1994aj} exploited the idea of S-duality, which expresses physics of strong-coupling phenomena in terms of the weakly coupled light dyons, thus providing an exact description of low-energy properties of the theory for an arbitrary coupling constant.
This approach was described with the help of the algebraic curve, which gives the prepotential as an analytical function of the scalar field, as was discussed for the $SU(2)$ gauge group in \cite{Seiberg:1994rs,Seiberg:1994aj}. The idea was extended to cover pure gauge theory with other gauge groups, gauge theory with matter, as well as used to study a number of related new phenomena in the $\CN=2$ supersymmetric gauge theory 
[3-43].
For the classical gauge groups ($A,B,C,D$ series)
% $A_{n},B_{n},C_{n}$ and $D_{n}$ 
the curve, which describes the solution is widely believed to be hyperelliptic, though   \cite{Martinec:1995by} suggested the non-hyperelliptic description for all gauge groups, which is based on the analogy with the integrable systems. Exceptional groups ($G,F,E$) 
% $G_2,F_4,E_6,E_7$ and $E_8$ 
prove to be more complicated for an analysis, see discussion in \cite{Abolhasani:1996ik,Landsteiner:1996ut,Alishahiha:2003hj}. 

The prepotential derived from this analysis provides a way to establish the magnetic and electric charges of light dyons, which were presented explicitly for the simplest gauge groups, including
$SU(2)$ 
\cite{Seiberg:1994rs}, 
$SU(3)$ \cite{Klemm:1995wp}, 
%$SO(5)$ \cite{Danielsson:1995is}, 
$SU(4)$ and $G_2$ \cite{Hollowood:1997pp}.
Clearly, the charges of light dyons are very interesting by themselves, which inspires their study for a general gauge group. This issue was addressed in  \cite{Hollowood:1997pp}, which based analysis on   \cite{Martinec:1995by} that related the Seiberg-Witten solution to the spectral curve of an integrable system. The work \cite{Hollowood:1997pp} provided a general procedure to derive the charges of the dyons, though the expressions found were involving.

The present work considers light dyons using basic properties of the theory directly, avoiding references to the curve that governs the theory. It is known from \cite{Seiberg:1994rs}
that light dyons describe the explicit breaking of the $\CN=2$ supersymmetry down to $\CN=1$ supersymmetry. Therefore matching the known fundamental properties of $\CN=2$ and $\CN=1$ supersymmetric gauge theories one can extract information related to properties of light dyons. The work is divided into two parts. Sections \ref{N=2} - \ref{monopole} summarize basic properties of the supersymmetric $\CN=1,2$ gauge theories. Sections \ref{chiral} - \ref{conc} derive and discuss expressions for the charges of dyons.

\section{Supersymmetric $\CN=2$ gauge theories  }
\label{N=2}
The supersymmetric  $\CN=2$ gauge theory  includes the scalar field $A$, two chiral spinors $\psi$ and $\lambda$, where the latter represents the gaugino, and the gauge field $v_\mu$,  all in the adjoint representation of a gauge group, which is a simple Lie group $\mathsf{G}$ with an algebra $\mathsf{g}$ \cite{Sohnius:1985qm}. The energy of the scalar field turns zero provided this field has a coordinate independent value that lies in the Cartan subalagebra $\gc$ of the gauge algebra, $A\in \gc \subset \mathsf{g}$, and satisfies $\Re \,(A)\propto \Im\,(A)$.  Thus, the scalar field can develop an expectation value in the vacuum, which makes the vacuum state degenerate, the moduli space is given by $\gc$, and 
and the scalar field in the vacuum, $A\in \gc$, can be treated as an $r$-dimensional vector, $A\equiv (A_1,\dots,A_r)$.

The vacuum expectation value of the scalar field breaks the gauge symmetry spontaneously. Generically, the symmetry  is broken down to $r$ products of gauge $U(1)$, $G \rightarrow U(1)\times \cdots \times U(1)$, where $r$ is the rank of the algebra $\mathsf{g}$. There also 
remains unbroken a discrete group of gauge transformations, which comprises the Weyl group of $\mathsf{g}$, as discussed below. In the perturbation theory region this gauge breaking generates masses for all degrees of freedom, except for those that correspond to the $r$ unbroken $U(1)$ gauge symmetries. As a result, there are  $r$ massless gauge fields in the theory, which are similar to photons; each such photon $v_\mu$ is accompanied by the corresponding massless fields $A,\psi$ and $\lambda$, which all have no electric charges and belong to the Cartan subalgebra $\gc$. 
The low-energy properties of the theory are described by one function, the prepotential $\CF$, which is a holomorphic function of the scalar field $\CF=\CF (A)$, as was argued in \cite{Seiberg:1988ur}. For a strong scalar field the coupling constant is weak. In this case one can write the prepotential explicitly in a simple form
\begin{eqnarray}
	\CF(A)\simeq \frac{i}{8\pi}\,\sum_\alpha \,(\alpha \cdot A)^2 \ln \frac{(\alpha \cdot A)^2~ }{\Lambda^2}~.
	\label{FPTh}
\end{eqnarray}
Here summation runs over all roots $\alpha$ of the algebra $\mathsf{g}$, the dot-product refers to the usual scalar product in an $r$-dimensional space, and $\Lambda$ is the conventional cut-off parameter. The limit of strong scalar field implies $|A^2|\gg \Lambda^2$. The notation $V^2\equiv V\cdot V$ for any $r$-dimensional vector $V$ is used throughout. Calculating the sum in   (\ref{FPTh}) with the logarithmic accuracy, i.e. presuming that 
$\ln(\alpha\cdot A)^2 \approx  \ln A^2$, which can be done when the scalar field is not close to a wall of the Weyl camera, one writes
\begin{eqnarray}
	\CF(A)\approx \frac{i}{4\pi}\, \hv \,A^2\ln \frac{A^2}{\Lambda^2}~,
	\label{Flog}
\end{eqnarray}
where $\hv$ is the dual Coxeter number of the algebra $\mathsf{g}$.
For relevant properties of simple Lie algebras see \cite{Bourbaki:2002,Di-Francesco:1997,Slansky:1981yr}. 
\footnote{The dual Coxeter number  $\hv=\hv(\mathsf{g})$ of the algebra $\mathsf{g}$, the eigenvalue of the quadratic Casimir operator in the adjoint representation $C_2(\mathsf{g})$,
and the Dynkin index of the adjoint representation $\chi_\mathrm{adj}(\mathsf{g})$ are all related, $2\hv(\mathsf{g})=C_2(\mathsf{g})=\chi_\mathrm{adj}(\mathsf{g})$,  see e.g. \cite{Di-Francesco:1997}, Eqs.(13.128),(13.134).} 
Deriving   (\ref{Flog}) the following identity was used
\begin{eqnarray}
	\sum_{\alpha}\,\alpha_i\,\alpha_j\,=\,2\,\hv\,\delta_{ij}~.
	\label{ahv}
\end{eqnarray}
Here and below the subscripts $i,j=1,\,\dots\,r$ refer to the Cartesian components of  $r$-dimensional vectors.   Equation (\ref{ahv}) is valid provided the roots are normalized conventionally, with large roots satisfying $\alpha^2=2$.

In \cite{Seiberg:1994rs} it is explained that the dual field $\AD$, which is defined by
\begin{eqnarray}
\AD=\frac{\partial \CF(A)} {\partial A}~,
	\label{AD}
\end{eqnarray}
plays a major role in the problem. Similarly to $A$, this dual field is an $r$-dimensional vector. Since $\CF$ is holomorphic, the dual field is
a holomorphic function of $A$ as well. In the weak coupling limit 
(\ref{FPTh}) this function reads
\begin{eqnarray}
A_\mathrm{D} \simeq \frac{i}{4\pi}\,\sum_\alpha \,\alpha \,(\alpha \cdot A) 
\left( \ln \frac{(\alpha \cdot A)^2}{\Lambda^2}+1\right) \approx 
 \frac{i}{2\pi} \, \hv\,A \, \ln \frac{A^2}{\Lambda^2}~,	
	\label{APTh}
\end{eqnarray}
where the last equality is written in the large-logarithm approximation introduced in   (\ref{Flog}). Effective coupling constants, which govern low-energy properties of the theory, are represented by a $r\times r$ matrix $\tau$, 
\begin{eqnarray}
	\tau_{ij}\,\equiv \,
	\left(\frac{\theta}{2\pi} +4\pi i\, g^{-2}\right)_{ij}\,=\,
	\frac{\partial {\AD}_i }{\partial A_j} ~.
	\label{tau}
\end{eqnarray}
Here $\theta$ and $g$ are the $r\times r$ matrices of theta-angles and proper coupling constants respectively.
  Equations (\ref{APTh}), (\ref{tau}) show that in the weak coupling region 
\begin{eqnarray}
	\tau_{ij}\approx \frac{i}{2\pi}\, \hv \ln \frac{A^2}{\Lambda^2}\,\,\delta_{ij}~.
	\label{tauPTh}
\end{eqnarray}
This equality implies that for the weak coupling there is only one coupling constant, whose asymptotic behavior is governed by the coefficient $b$ of 
the Gell-Mann - Low beta-function that equals 
\begin{eqnarray}
	b=2 \hv\,, 
	\label{b=2}
\end{eqnarray}
in accord with expectations for the $\CN=2$ supersymmetric gauge theory, see e.g.  \cite{Peskin:1997qi}.

The spontaneous breaking of the gauge symmetry keeps intact a set of discrete gauge transformations, the Weyl group $\mathsf{W}$ of the algebra $\mathsf{g}$, which comprises reflections in  hyperplanes orthogonal to roots. Such a reflection transforms any $r$-dimensional vector $V$ via $V\rightarrow V'=\rho_\alpha V$, where $\rho_\alpha$ is an $r\times r$ matrix
\begin{eqnarray}
	\rho_\alpha\,=\,
	1-\alpha \,\otimes\,\av~,
	\label{R}
\end{eqnarray}
Here $\alpha^\vee$ indicates a coroot
\begin{eqnarray}
	\alpha^\vee=\,\frac{2 }{\,\alpha^2}\,\,\alpha
	\label{av}
\end{eqnarray}
and a conventional notation for $r\times r$ matrices $(1)_{ij}=\delta_{ij}$, $(\,\alpha \,\otimes \,\av )_{ij}= \alpha_{i}\,\av_{j}$ is employed throughout (being complemented 
below by an obvious $(0)_{ij}=0$).  It is convenient \cite{Seiberg:1994rs} to introduce the $2r$-dimensional vector 

\begin{equation}
	\varPhi\,=\,\begin{pmatrix} ~\AD \\ A \end{pmatrix}~.
	\label{Phi}
\end{equation}  
Under the transformation of $A$ and $\AD$ by $\rho_\alpha \in \mathsf{W}$ this vector is transformed according to
\begin{align}
&\Phi
 \rightarrow  \Phi' \,  = \,P_\alpha \,\Phi~,
\label{Rcl}
\\ 
& P_{\,\alpha}\,=\,\begin{pmatrix} \rho_\alpha &0 \\ 0 & \rho_\alpha \end{pmatrix}~.
\label{Rrho}	
\end{align}
All entries in the $2\times 2$ block matrix here are $r\times r$ matrices, similar notation is used below. 

However, if one considers a continuous transformation of the scalar field, which starts from $A$ and ends at $\rho_\alpha A$, then   (\ref{APTh}) shows that there arises an additional contribution to $\AD$, which can be interpreted as a quantum correction. It is proportional to the variation of the logarithmic function that acquires $\pm\,i\pi$ when the scalar field crosses the wall of the Weyl camera, which is orthogonal to $\alpha$.
The sign of this variation depends on the way the crossing is fulfilled. 
Taking this variation as $-i\pi$ (we return to this point considering   (\ref{NewR}) below) one can write the monodromy, which was suggested in \cite{Danielsson:1995is}, that describes the transformation of the scalar field along the path considered 
\begin{eqnarray}
	\Phi\,
\rightarrow  \,\Phi' \,=\,
{R}_{\,\alpha}\,\Phi~,
	\label{Rq}
\end{eqnarray} 
where ${R}_\alpha$ is
\begin{align}
{R}_\alpha &\,=\,\begin{pmatrix} \rho_\alpha & \alpha \otimes \alpha \\ 0 & \rho_\alpha \end{pmatrix}\,=\,P_\alpha\,T_\alpha~.
\label{Ra}
\end{align}
Here $P_\alpha$ is the Weyl reflection   (\ref{Rrho}) and $T_\alpha$ is the matrix
\begin{equation}
T_\alpha \,=\,
\begin{pmatrix} 1 & -\alpha \otimes \alpha \\ 0 & ~~~1 \end{pmatrix}~.
\label{Ta}
\end{equation}
It was shown in \cite{Danielsson:1995is} that the set of $r$ matrices $R_\alpha$ defined in   (\ref{Ra}) generates the Brieskorn braid group.

There are global symmetries important for the Seiberg-Witten solution.
One of them represents the global $SU(2)$ symmetry, which is specific to the $\CN=2$  supersymmetry. Transformations from this $SU(2)$ group treat $\psi$ and $\lambda$ as a doublet, leaving the scalar and vector fields $A,v_\mu$ invariant. There is also a chiral
$U(1)$ symmetry, which on the classical level manifests itself via the following transformations of the fields
\begin{eqnarray}
\label{U(1)}  
	& \vartheta \rightarrow e^{i\gamma} \vartheta~,\quad
	\psi  \rightarrow e^{i\gamma}\psi~,\quad
	\lambda \rightarrow e^{i\gamma} \lambda~,&
	\\  
	\label{2gammaA}
	& A \rightarrow e^{2i\gamma} A~.
\end{eqnarray}
Here $\vartheta$ is a conventional anti-commuting variable of the $\CN=1$ superspace \cite{Wess:1991}. Quantum corrections break this symmetry to $Z_{\,4\hv}$, forcing the phase $\gamma$ in   (\ref{U(1)}), (\ref{2gammaA}) to take only discrete values
\begin{eqnarray}
	\gamma=2\pi\,\frac {m} {4\hv}~,\quad m=0,1,\,\dots\,,4 \hv-1~.
	\label{gamma}
\end{eqnarray}
The effects, which lead to this restriction on $\gamma$ can be attributed to the variation of the $\theta$-angle of the theory, which takes place due to the chiral transformation   (\ref{U(1)}). This variation reads $\Delta \,\theta \,=\,4\hv\gamma$. The symmetry persists provided this variation is an integer of $2\pi$, 
\begin{equation}
\Delta \,\theta\,=\,4\hv\gamma\,=\,2\pi m~, 	
	\label{Delta}
\end{equation}
which leads to   (\ref{gamma}).
Overall, the global symmetry is $\left( SU(2)\times {Z}_{\,4\hv}\right)/{Z}_2$, the divisor ${Z}_2$ eliminates double-counting of the center of $SU(2)$, which is also present in ${Z}_{\,4\hv}$.

The transformation of the scalar field $A$ in   (\ref{2gammaA}) is accompanied by the transformation of the dual field $\AD$. Using   (\ref{APTh}) one finds 
\begin{eqnarray}
\AD\rightarrow \AD'= \exp(\,2i\gamma\,) \left(\AD-\frac{\gamma}{\pi}\,\sum_\alpha\,(\alpha\cdot A)\,\alpha \right)	=
\exp\left( \,\pi i/\hv\, \right) \left(\AD-A \right)\,.	
	\label{ADZ}
\end{eqnarray}
Here the second term in the big parentheses originates from the logarithmic function in 
  (\ref{APTh}). The last identity in   (\ref{ADZ}) is written using   (\ref{ahv}) and assuming $m=1$ in    (\ref{gamma}).
  Equation (\ref{ADZ}) shows that the defining element of the chiral ${Z}_{\,4\hv}$ symmetry  manifests itself via the following transformation of the scalar field 
\begin{eqnarray}
\Phi\,\rightarrow \,
\Phi' \, = \,
\exp\left( \,\pi i/\hv \right) M\,\Phi~,
	\label{ADgamma}
\end{eqnarray}
where the $2r\times 2r$ matrix $M$ reads
\begin{equation}
M\,=\,\begin{pmatrix} 1 & \!-1 \\ 0 & \,~1 \end{pmatrix}~.	
\label{M}
\end{equation}
This transformation can be considered as a monodromy that arises when the phase $\gamma$  is treated as a continuous variable that varies from $\gamma=0$ to the value $\gamma=2\pi/\hv$ allowed by   (\ref{gamma}). 

Explicit forms for the monodromies in  (\ref{Rq}) and (\ref{ADgamma}) were presented in the weak coupling limit. However, the prepotential, and therefore the dual field remain holomorphic functions of the scalar field even in the strong-coupling region. Consequently, a continuous variation of the scalar field is accompanied by a continuous variation of the dual field (provided no singularities are crossed). This implies that the discrete-valued matrices on the right-hand sides of  (\ref{Rq}),(\ref{ADgamma}) do not change, if we 
consider both these transformations as monodromies under continuous variations of the scalar field and the path along which the transformation is defined. Thus, the transformations in  (\ref{Rq}),(\ref{ADgamma}) remain well defined even when the strong-coupling region is considered.

\section{ Supersymmetric $\CN=1$ gauge theory}
\label{N=1}
The supersymmetric $\CN=1$ gauge theory includes the gauge  field  $ v_\mu $ and gaugino $ \lambda $ \cite{Wess:1991}.  On the classical level the theory possesses the global chiral $U(1)$ symmetry
\begin{eqnarray}
 \vartheta\rightarrow e^{i\delta}\,\vartheta~,\quad 	
 \lambda \rightarrow e^{i\delta}\,\lambda~. 	
	\label{chiralN1}
\end{eqnarray}
Quantum corrections break the chiral symmetry to ${Z}_{\,2\hv}$. The latter  group is defined by transformations in (\ref{chiralN1}), in which the phase takes the following discrete values
\begin{eqnarray}
	\delta\,=\,2\pi\,\frac{k} {2\hv}~,\quad k=0,1,\,\dots\,2\hv-1~.
		\label{delta}
\end{eqnarray}
The resulting ${Z}_{\,2\hv}$ is further broken spontaneously down to ${Z}_{\,\hv}$. Overall, the breaking of the chiral symmetry follows the pattern 
\begin{eqnarray}
	U(1)\,\rightarrow \, {Z}_{\,2\hv}\,
	\rightarrow \, {Z}_{\,\hv}~.
	\label{chiralN=1}
\end{eqnarray}
This pattern manifests itself through the gaugino condensate $\langle\lambda\lambda\, \rangle$, which is present in the vacuum, 
$\langle\lambda\lambda\, \rangle\ne0$, see e.g.  \cite{Peskin:1997qi,Shifman:1999mv} for  discussion and references.
%Shifman:1987ia 
%Morozov:1987hy
According to   (\ref{delta}), (\ref{chiralN=1}) 
the ${Z}_{\,\hv}$ global symmetry results in a transformation of the gaugino condensate 
\begin{eqnarray}
\langle\lambda\lambda\, \rangle\,\rightarrow \,\exp\left(\, 2\pi i /\hv \right)\,\langle\lambda\lambda\, \rangle~.
	\label{ll}
\end{eqnarray}
It is  commonly believed that the symmetry breaking (\ref{chiralN=1}) and the presence of the gaugino condensate generate a mass gap. It is also presumed that there is no vacuum degeneracy, except for the one specified in   (\ref{ll}), which makes the phase of the gaugino condensate a convenient marker for the vacuum. 
Equation (\ref{ll}) shows that  shifts of this phase by $2\pi k/\hv,~k=0,1,\,\dots \,\hv-1$
constitute ${Z}_{\,\hv}$. Different values of the phase mark different vacua. The
vacuum reveals therefore the $\hv$-fold degeneracy. This assessment complies with calculations based on the Witten index $I_\mathrm{\,W}$ \cite{Witten:1982df}, which counts the difference between the number of bosonic $n_\mathrm{b}$ and fermionic $n_\mathrm{f}$ zero-energy states
\begin{eqnarray}
I_\mathrm{\,W}=n_\mathrm{b}-n_\mathrm{f}~,
	\label{IW}
\end{eqnarray}
and is designed to be invariant under continuum transformations of parameters of the theory. The calculations of \cite{Witten:1982df,Witten:1997bs} based on this property of the index showed that for the supersymmetric $\CN=1$ gauge theory the index reads
\begin{eqnarray}
I_\mathrm{\,W}\,=\,\hv~.	
	\label{I=h}
\end{eqnarray}
For the pure gauge theory it is presumed that $n_\mathrm{f}=0$. Equation (\ref{I=h}) implies therefore that there are precisely $\hv$ vacua, in accord with the symmetry breaking pattern (\ref{chiralN=1}) and chiral transformations of the gaugino condensate (\ref{ll}).

\section{Monopoles and dyons}
\label{monopole}
The Seiberg-Witten solution expresses the low-energy properties of the $\CN=2$ supersymmetric theory at strong coupling in terms of light monopoles and dyons. The magnetic $g$ and electric $q$ charges of a dyon describe its interaction with $r$ different electro-magnetic fields that are present in the theory. Therefore both $g$ and $q$ are described by $r$-dimensional vectors. Moreover, the electric and magnetic charges lie in the lattice  of roots and coroots of the algebra $\mathsf{g}$ respectively, see
the review \cite{Weinberg:2006rq},
\begin{align}
\label{qdirect}
	q &\,=\,\sum_{\alpha \,\in\,\, \Delta} \,m_\alpha\,\alpha~,\quad  m_\alpha \in {Z}~, \\
	\label{gdual}
	g &\,=\,\sum_{\alpha \,\in \,\,\Delta}  \,n_\alpha\,\alpha^\vee~,\quad  n_\alpha \in {Z}~.
\end{align}
Equation (\ref{qdirect}) complies with the fact that in the pure gauge theory
all fields are in the adjoint representation of $\mathsf{g}$, which is associated with the lattice of roots. From   (\ref{qdirect}) one derives   (\ref{gdual}) by applying the Dirac-Schwinger-Zwanziger quantization condition, which states that  
magnetic and electric charges  $g_1,~q_1$ and $g_2,~q_2$  of any two dyons satisfy 
\begin{equation}
g_1\cdot q_2-g_2\cdot q_1\,\in \,{Z}~, 	
	\label{dsz}
\end{equation}
which can be conveniently casted into
\begin{eqnarray}
	\CG_1\, \Omega\,\,\CG_{\,2}^{\,T} \,\equiv \,(\,g_1,\,q_1\,) \,
	\begin{pmatrix} ~~0 & 1~ \\ -1 & 0~ \end{pmatrix}\,
	\begin{pmatrix} g_2 \\ q_2 \end{pmatrix}\,
	\in {Z}~.
	\label{DSZ}
\end{eqnarray}
Here 
\begin{equation}
	\CG=(\,g,\,q\,)
	\label{Ggq}
\end{equation}
represents magnetic and electric charges of a dyon, and the identity in   (\ref{DSZ}) employs the symplectic metric $\Omega$
\begin{equation}
	\Omega\,=\,\begin{pmatrix} ~~0 & 1~ \\ -1 & 0~ \end{pmatrix}~.
	\label{Omega}
\end{equation}
Equations (\ref{qdirect}),(\ref{gdual}) comply with (\ref{DSZ}) since any two roots $\alpha,\beta \in \mathsf{g}$ satisfy $\alpha^\vee\cdot\beta \in {Z}\,$. 
It suffices to limit summation in   (\ref{qdirect}),(\ref{gdual}) to simple roots, which is specified as $\alpha\in \Delta$, where $\Delta$ is a set of $r$ simple roots of $\mathsf{g}$.

The dyons in the Seiberg-Witten solution are presumed to be BPS states \cite{Bogomolny:1975de,Prasad:1975kr}. Consequently, 
as was noted in \cite{Witten:1978mh}, the mass $m_{\CG}$ of a dyon with the charge $\CG=(g,q)$ is related to the central charge $\CZ_{\,\CG}$
\begin{align}
\label{MZ}
	 m_{\,\CG}   & \,=\,  2^{\,1/2} \,\,|\,\CZ_{\,\CG}\,|~, \\ 
\label{Z}	 
	 \CZ_{\,\CG}     &   \,=\,   g \cdot \AD+q\cdot A \,\equiv \,\CG \,\varPhi~.
	\end{align}

\section{Chiral transformations for dyon charges}
\label {chiral}
Our goal is to study { light} dyons, i.e. those dyons that become massless at an appropriate value of the vacuum expectation value of the scalar field. In the vicinity of this value the mass of the light dyon is small. When a dyon becomes massless, one can consider breaking the $\CN=2$ supersymmetry to the $\CN=1$ case  by introducing a mass term for the fields $A$ and $\psi$ that comprise the $\CN=1$ chiral multiplet $\Phi$. As a result the vacuum of the $\CN=1$ supersymmetric theory arises, in which a condensate of dyons is developed \cite{Seiberg:1994rs}. 
This phenomenon has an important impact on the problem by making all fields massive. Thus, the dyons, which are initially created within the $\CN=2$ supersymmetry can explain the origin of the mass gap for the $\CN=1$ theory. The light dyons become massless in the strong-coupling region of the $\CN=2$ theory, but the S-duality implies that their behavior is governed by the dual coupling constant, which is weak in this region, making theoretical framework reliable. 

These arguments due to Seiberg and Witten \cite{Seiberg:1994rs} show that massless dyons bridge properties of the $\CN=2$ and $\CN=1$ supersymmetric gauge theories. 
This phenomenon, which was discussed in detail in  \cite{Seiberg:1994rs}  for an example of the $SU(2)$ gauge group, will be considered in the present work as a general feature of the Seiberg-Witten solution valid for an arbitrary gauge group.
Let us use this property of the problem as an opportunity to study dyons by applying known properties of $\CN=2$ and $\CN=1$ supersymmetric theories. Take some light dyon with the charge $\CG$ defined by   (\ref{Ggq}). Assume that a condensate of dyons is developed, 
in which this particular dyon is contributing when the $\CN=2$ supersymmetry is broken to 
the $\CN=1$ case (we will argue below that several dyons necessarily condense together). The resulting condensate of dyons describes some vacuum of the $\CN=1$ theory, call this vacuum state $|\,0\, \rangle$. Clearly this vacuum is not unique,   (\ref{I=h}) shows that there are other vacua, $\hv$ of them overall. Let us assume we make a transition from $|\,0\, \rangle$ into some other vacuum $|\,0\,'\, \rangle$, 
$|\,0\, \rangle\rightarrow |\,0\,'\, \rangle$. Within the framework of $\CN=1$ supersymmetry this transition can be associated with a discrete chiral transformation in   (\ref{ll}). 

Consider now the same transformation $|\,0\, \rangle\rightarrow |\,0\,' \rangle$ in terms of dyons. The vacuum $|\,0\, \rangle$ was presumed to arise due to condensation of dyons having charge $\CG$. Similarly, we should expect that a different vacuum state
$|\,0\,' \rangle$ is created due to condensation of dyons with some different charge $\CG\,'$. In other words, a transition from one vacuum state to another, $|\,0\, \rangle\rightarrow |0\,' \rangle$, should be accompanied by a transformation of the dyon charge, from the initial charge $\CG$ to the different one $\CG\,'$,  $\CG \rightarrow\CG\,'$. We know that transitions between different vacua $|\,0\, \rangle\rightarrow |\,0\,' \rangle$ in the $\CN=1$ theory are classified by the symmetry group $Z_{\,\hv}$ that arises from the discrete chiral transformations, as per  (\ref{chiralN=1}),(\ref{ll}). Consequently, we are to expect that a similar symmetry group should classify related transformations $\CG \rightarrow\CG\,'$ of the dyon charges.

Let us specify this symmetry group. Since the dyons are created within the framework of $\CN=2$ supersymmetry, we need to look at discrete symmetries available in the $\CN=2$ case.  The transitions between different vacua of the $\CN=1$ theory are generated by the chiral transformations. One has to expect therefore that similar chiral transformations are responsible for the transformations between different charges of dyons. Thus, there should exist a connection between the discrete $Z_{\,\hv}$ symmetry, which governs transformations between different vacua of the $\CN=1$ supersymmetric theory, and the chiral $Z_{\,4\hv}$ symmetry of the $\CN=2$ supersymmetry. 

The action of the $Z_{\,4\hv}$ symmetry on the charge of a dyon can be derived from   (\ref{Z}) by re-interpreting the chiral transformation of the scalar field in this equation in terms of a transformation of the dyon charge, see   (\ref{G->GM}) below. 
  Equation (\ref{2gammaA}) states that the scalar field has charge 2 under chiral transformations.
We conclude that the influence of the chiral group $Z_{\,4\hv}$ on the 
charge of a dyon can manifest itself only through a smaller group $Z_{\,2\hv}$. 
This latter group matches perfectly the full discrete chiral group $Z_{\,2\hv}$ of the $\CN=1$ supersymmetry. Thus, we observe a correspondence between chiral transformations of the dyon charges within the $\CN=2$ supersymmetry and the discrete chiral symmetry of the $\CN=1$ theory.

The vacuum of $\CN=1$ theory breaks the chiral group $Z_{\,2\hv}$ spontaneously to $Z_{\,\hv}$, see  (\ref{chiralN=1}),(\ref{ll}). This breaking effectively eliminates the difference between the fields $\lambda$ and $-\lambda$ in the sense that both these fields give the same contribution to the gaugino condensate $\langle  \lambda \lambda\rangle$ in the vacuum.
For future reference let us cast this statement, i.e. the fact that the fields 
$\lambda$ and $-\lambda$ give the same contribution to the gaugino condensate, into
\begin{equation}
	\lambda\,\equiv -\lambda~.
	\label{la-la}
\end{equation}
In other words, a spontaneous breaking $Z_{\,2\hv}\rightarrow Z_{\,\hv}$ takes place because  the elements $1$ and  $-1$ of the group $Z_{\,2\hv}$, which is taken in the multiplicative notation, are identified modulo $Z_{\,2}$ 
\begin{equation}
Z_{\,\hv}\, =\, Z_{\,2\hv}/Z_{\,2}~.
	\label{1-1}
\end{equation}
Since we presume that the structure of the $\CN=1$ theory is reproduced by the condensates of light dyons, we have to admit also that the pattern of the symmetry breaking described by   (\ref{1-1}) should govern the behavior of condensates of light dyons as well. This means that the identity $1\,\equiv -1$, which is acknowledged in   (\ref{1-1}) for vacua of the $\CN=1$ theory, should be applicable for the condensates of light dyons as well. 
The properties of light dyons are governed by the scalar field, thus prompting this identity to be implemented in terms of the scalar field. 
Since $\lambda$ and $A$ belong to the same $\CN=2$ hypermultiplet, 
there exits only one option, namely we have to admit that  (\ref{la-la}),(\ref{1-1}), when applied within the framework of the $\CN=2$ supersymmetric theory imply that the two values of the scalar field, $A$ and $-A$, should be identical
\begin{equation}
A\equiv -A~.
	\label{A-A}
\end{equation}
This identity should govern the behavior of light dyons. Suppose that the field $A$ makes some dyons massless. Consider the condensate of these dyons, which is related to some vacuum of the $\CN=1$ supersymmetric gauge theory, when the $\CN=2$ supersymmetric gauge theory is broken to the $\CN=1$ case. Then   (\ref{A-A}) is to be understood as a statement that the field $-A$ also produces massless dyons, and the condensate created by these dyons is identical to the condensate created when the value of the scalar field equals $A$. The identity between the two condensates leads to the identity of the related two vacua of the $\CN=1$ theory, in accord with   (\ref{la-la}).
Thus,   (\ref{la-la}) and (\ref{A-A}) present two different points of view on the same spontaneous breaking of symmetry. Equation (\ref{la-la}) describes it in terms of the vacuum state of the $\CN=1$ supersymmetric gauge theory, while (\ref{A-A}) looks at it from the perspective of the $\CN=2$ supersymmetry, where the dyons are defined.

From   (\ref{A-A}) one deduces that when the condensate of dyons arises then the group $Z_{\,2\hv}$, which governs chiral transformations of the scalar field, is broken spontaneously to $Z_{\,\hv}$, in accord with   (\ref{1-1}). This discussion can be summed up as the following pattern  of the symmetry breaking
\begin{equation}
	U(1)\rightarrow Z_{\,4\hv}\rightarrow Z_{\,2\hv}\rightarrow Z_{\,\hv}~,
	\label{chiralN=2}
\end{equation}
which describes how the chiral symmetry of the $\CN=2$ gauge theory is implemented for light dyons. The first step here shows a conventional breaking of the classical chiral symmetry by quantum effects, which was mentioned  in Section \ref{N=2}. The second step is justifies by the fact that the chiral transformations of the charges of dyons are related to the scalar field that has charge 2. The last, third step is valid for light dyons in view of   (\ref{A-A}). Note the resemblance between   (\ref{chiralN=2}) and (\ref{chiralN=1}). 

We conclude that the subgroup $Z_{\,\hv}$ of the chiral group, which describes transformations between different vacua for the $\CN=1$ supersymmetry, $|\,0\, \rangle \rightarrow |\,0\,'\, \rangle$, is matched in the $\CN=2$ supersymmetric theory by a similar subgroup $Z_{\,\hv}$, which describes transformations between the charges of dyons, and which also originates in the chiral symmetry.

Let us calculate charges of dyons in different condensates, which comprise different vacua of the $\CN=1$ gauge theory. The discussion above shows that it suffices to apply chiral transformations of the $\CN=2$ supersymmetry to the charge of one light dyon. Consider the transformation of the mass $ m_{\,\CG }$ of a dyon under the chiral transformation of the scalar field. From    (\ref{ADgamma}) one deduces
\begin{equation}
	m_{\,\CG}\,=\,2^{\,1/2} \,\,\left| \,\,\CG\,\Phi\,\right| \rightarrow \,m_{\,\CG\,'}\, =\,
	2^{\,1/2} \,\,
	\left|\,\,\CG\,\Phi\,'\,\right|\,=\,
	2^{\,1/2} \,\,
	\left|\,\,\CG\, M \,\Phi\,\right|~.
	\label{mm}
\end{equation}
We can now interpret (\ref{mm}) as the definition of the chiral transformation for the charge of the light dyon, thus deriving
\begin{equation}
	\CG\, \rightarrow \, \CG\,'\,=\,\CG\,M~,
	\label{G->GM}
\end{equation}	
or more explicitly 
\begin{equation}
	(g,q) \rightarrow (g',q')\,=\,(g,q-g)~.
	\label{gqgq}
\end{equation}
This transformation complies with the Witten effect \cite{Witten:1979ey}, which states that the magnetic charge $g$ gives a contribution $\delta q$
to the electric charge  of a dyon
\begin{equation}
\delta q\,=\,-g\,\theta/(2\pi)~, 	
	\label{eq:}
\end{equation}
where $\theta$ is the theta-angle. The effect is often discussed for the pure electromagnetic theory, whereas in the case considered there are $r$  electromagnetic theories present simultaneously, but this minor generalization brings no complications since 
different electromagnetic fields do not interact. The chiral transformation, which changes the charge of a dyon, results in the variation of the $\theta$-angle by $2\pi$, see   (\ref{Delta}), forcing thus a change of the electric charge of $\delta q=-g$, in accord with   (\ref{gqgq}).

Applying the transformation (\ref{G->GM}) $m$ times to the dyon with the charge $\CG$, and taking into account also inversed transformations we find a set of dyons with charges 
\begin{subequations}
\label{Gm}
\begin{align}
\label{Mm} 
\CG\,'\,&=\,\CG\,M^{\,m}~,\\
\label{qm}
(\,g\,'\,,\,q\,'\,)\,& =\,(\,g,\,q-m\,g\,)~.		
\end{align}
\end{subequations}
Consider a restriction on the integer $m$ here that stems from   (\ref{A-A}).
Equation (\ref{2gammaA}) shows that in order to fulfill the transformation $A \rightarrow -A$ 
the integer $m$ in   (\ref{gamma}), and consequently in (\ref{Mm}),(\ref{qm}), should satisfy $m=\hv$.
According to   (\ref{A-A}) the transformation $A \rightarrow -A$
does not change the vacuum of the $\CN=1$ supersymmetric gauge theory.
In other words, a variation of $m$ in (\ref{Mm}),(\ref{qm})
by $\hv$ keeps the vacuum of the $\CN=1$ theory intact.
We conclude that in order to describe different vacua of the $\CN=1$ theory it suffices to take  $m$ in  (\ref{Mm}),(\ref{qm}) modulo $\hv$
\begin{equation}
m \in Z_{\,\hv}~. 	
	\label{mZh}
\end{equation}
As an illustration of the validity of   (\ref{Gm}), (\ref{mZh}) recall the $SU(2)$ gauge group \cite{Seiberg:1994rs}, when two dyons with the charges
\begin{eqnarray}
	\frac{\CG_0}{2^{\,1/2}}\,=\,(1,0)~,\quad  \frac{\CG_{1}}{2^{\,1/2}}\,=(1,-1)
		\label{SU2}
\end{eqnarray}
play a major role. Note that $\mathsf{G}=SU(2)$ implies $r=1$, $\alpha=2^{\,1/2}$, $\hv =2$, and $Z_{\,\hv}=Z_{\,2}$, and that
notation in   (\ref{SU2}) follows definitions given in (\ref{qdirect}),(\ref{gdual}) and (\ref{Ggq}), which prompt the coefficient $2^{\,1/2}=\alpha$. One observes that the necessity of having two dyons, as well as values of their charges in   (\ref{SU2}) for $\mathsf{G}\,=SU(2)$ comply with  (\ref{Gm}),(\ref{mZh}).

\section{Minimal set of light dyons}
\label{minimal}
Let us find a minimal set of light dyons, including in this set only those dyons that are absolutely necessary to describe any possible condensate of dyons, which is created when the $\CN=2$ supersymmetry is broken down to the $\CN =1$ case. 
To construct this minimal set start from the  simplest dyon, a monopole. Thus presume that there exists a light monopole with the charge   $\CG= (\av,0)$, $\alpha \in \Delta$. In this case   (\ref{qm}) shows that there exist a series of light dyons with charges $\CG_{\,\alpha, \,m}$,
\begin{equation}
	\CG_{\,\alpha, \,m} \, = \,(\,\av,-m \,\av )~,\quad m\,\in\,Z_{\,\hv}~,
	\label{Gam}
\end{equation}
where $\av$ is a fixed simple coroot. The monopole charge is reproduced here when $m=0$.

Assume now that the monopole with the charge $\CG_{\,\alpha, 0}$ participates in the creation of a condensate, which arises when the $\CN=2$ supersymmetry is broken down to $\CN=1$ supersymmetry. According to \cite{Seiberg:1994rs} the condensation of dyons is responsible for creation of a vacuum of the $\CN=1$ supersymmetric gauge theory. 
It was shown in Section \ref{chiral} that other dyons with charges given in   (\ref{Gam}) arise due to chiral transformations. Therefore, each such chiral transformation brings a condensate of monopoles into some other condensate, which is created with the help of a dyon with the charge $\CG_{\,\alpha, \,m}$.  Overall, as   (\ref{mZh}) states, there are $\hv$ different dyons described by the series (\ref{Gam}). 
Correspondingly, there are $\hv$ different condensates of dyons, which in turn correspond to $\hv$ different vacua of the $\CN=1$ supersymmetric gauge theory. 

The existence of the mass gap in the vacuum of the $\CN=1$ supersymmetric gauge theory puts further restriction on the condensate of dyons. Consider a light dyon with the charge $\CG_{\,\alpha, \,m}$. Consider further a condensate, in which this dyon plays a role. For the sake of an argument presume firstly that this condensate includes only dyons with the charge $\CG_{\,\alpha, \,m}$, i.e. there are no other dyons with different charges in this condensate. However, this presumption runs into contradiction. 
The dyons of charge $\CG_{\,\alpha, \,m}$ interact  with only those degrees of freedom, `dual photons' and their superpatners, that originate from the gauge $U(1)$-group specified by the vector $\alpha$ that defines the charge $\CG_{\,\alpha, \,m}$ of the dyon. All other degrees of freedom, which are related to other $r-1$ available gauge $U(1)$-groups, do not interact with these dyons. Consequently a hypothetical condensate constructed from the dyons 
with the charge $\CG_{\,\alpha, \,m}$
is not able to break these $r-1$ gauge  $U(1)$ symmetries, thus leaving the 
corresponding  $r-1$ degrees of freedom massless. This contradicts the existence of the mass gap in the $\CN=1$ supersymmetric gauge theory. 

To remedy this problem one has to admit that the proper condensate of dyons,
which describes the breaking of the supersymmetry $\CN=2\rightarrow \CN=1$,
is created by several dyons with different charges, which become massless in the $\CN=2$ gauge theory simultaneously and which simultaneously go into the condensate state, when the $\CN=2$ supersymmetry is broken, $\CN=2\rightarrow \CN=1$ . Then any given degree of freedom among the $r$ available  gauge $U(1)$-symmetries would interact with some dyon in this condensate. As a result, this condensate is able to break all $r$ dual gauge $U(1)$-symmetries producing the mass gap, in accord with properties of the supersymmetric $\CN=1$ gauge theory.

We see that there should exist $r$ different massless dyons, which participate in the creation of the necessary condensate. Let us verify that the condensate is produced by $r$ dyons, which have charges
\begin{equation}
\CG_{\,\alpha, \,m} \, = \,(\,\av,-m \,\av )~,\quad \alpha \, \in \,\Delta~,
	\label{aid}
\end{equation}
where $m$ is fixed. A first test for this assessment gives the Dirac-Schwinger-Zwanziger quantization condition. The condensate constructed from a set of $r$ dyons can be described by conventional methods only if the dyons are mutually local 
\cite{t'Hooft:1981ht}. This means that the quantization condition    (\ref{DSZ}) for any two dyons from this set should read
\begin{equation}
		\CG_1\, \Omega\,\,\CG_{\,2}^{\,T} \,=\,0~.
	\label{zero}
\end{equation}
The charges in   (\ref{aid}) clearly satisfy this condition
\begin{equation}
\CG_{\,\alpha, \,m} \,\Omega ~\,\CG_{\,\beta, \,m} ^{\,T} =0~, \quad\quad\alpha,\beta\,\in\,\,\Delta~.	
	\label{ab}
\end{equation}
To compare   (\ref{aid}), (\ref{zero}) with the known results remember that according to \cite{Klemm:1995wp}, which studied the case of $\mathsf{G}=SU(3)$, two mutually local dyons can become massless simultaneously, in agreement with (\ref{aid}) and (\ref{zero}) since $\mathsf{G}=SU(3)$ implies $\hv=2$. We will return to this comparison in more detail in Section \ref{comp}. Condition (\ref{zero}) specifies mutually local light dyons. There are known particular situations, which arise near the cusps, when several mutually nonlocal dyons become massless simultaneously, as was discussed in \cite{Argyres:1995jj}, though these cases will remain outside the scope of the present work.

One more important test for the fact that   (\ref{aid}) gives charges of those light dyons that create the condensate when the $\CN=2$ supersymmetry is broken, $\CN=2\rightarrow\CN=1$, is provided by the Weyl symmetry. Equations (\ref{R}), (\ref{Rcl}),(\ref{Rrho}) define the Weyl reflection for the scalar field. Combining them with   (\ref{MZ}),(\ref{Z}) for the mass of dyons, one defines the Weyl reflection for the charges of dyons.  Thus, one finds that the Weyl reflection for the charge $\CG$ has an expected form
\begin{equation}
	\CG\,\rightarrow \,\CG'\,=\,\CG \,P_\beta~,
	\label{GG'}
\end{equation}
where the reflection $P_\beta$ is taken in the hyperplane orthogonal to the root $\beta$.
Applying the transformation (\ref{GG'}) to the charge $\CG=\CG_{\,\alpha, \,m}$ one derives from (\ref{GG'}),(\ref{Rrho}),(\ref{R})
\begin{align}
& \CG_{\,\alpha, \,m}~\rightarrow ~\CG_{\,\alpha, \,m}\,P_\beta\,=\, \CG_{\,\alpha',\,m}	~,\\
& \alpha' \,=\,\rho_{\beta}\,\alpha \, =\, \alpha-\beta\,\,(\,\beta^\vee \cdot \alpha\,)~.
	\label{GGrho}
\end{align}
Several Weyl reflections allow one to transform the charge $\CG_{\,\alpha, \,m}$ into any other charge $\CG_{\,\alpha',\,m}$, $\alpha'\in \Delta$, provided the lengths of roots $\alpha$ and $\alpha'$ are same, $\alpha'{\,^2}=\alpha^2$. It follows from this that if the dyon with the charge $\CG_{\,\alpha, \,m}$ is light, i.e. it becomes massless for some value of the scalar field, then all other dyons with the charges $\CG_{\,\alpha',\,m}$, $\alpha'\in \Delta$, $\alpha'{\,^2}=\alpha^2$, are also light, i.e. become massless for some values of the scalar field. In general case these values of the scalar field are different for different light dyons. 

For simply-laced $\mathsf{g}$ this discussion shows that all dyons with charges specified by   (\ref{aid}) are light ones. Moreover, this statement remains valid for nonsimply-laced  $\mathsf{g}$ as well.  Assume that there is one massless dyon, whose charge satisfies   (\ref{aid}) and equals $\CG_{\alpha,\,m}$. Assume, for example, that the simple root $\alpha$ is long, $\alpha^2=2$. Then we already know that there exist also massless dyons with charges $\CG_{\alpha',\,m}$,
where $\alpha'$ are simple large roots, $\alpha'\in\Delta$, $\alpha'^2=2$. We also know that the total number of massless dyons should be $r$. Therefore there must exist additional dyons.
We can assume that their magnetic charges equal $\beta^\vee$, where $\beta$ is any simple short root, $\beta\in\Delta$, $\beta^2<2$. Let us call the electric charges of these dyons $q_\beta$. Thus the total charges of the additional dyons are presumed to be $\CG_\beta=(\beta^\vee,q_\beta)$. Let us verify that these charges comply with   (\ref{aid}), which allows all these dyons to participate in the condensate that is responsible for the $\CN=2\rightarrow \CN=1$ supersymmetry breaking.
In other words let us verify that $\CG_\beta=\CG_{\beta,\,m}$.
Consider the quantization conditions   (\ref{zero}) for all available $r$ massless dyons, which read 
\begin{align}
&\av\cdot(q_\beta+m \beta^\vee)\,=\,0~,
\label{bb}
\\
&\beta^\vee\cdot q_\gamma-\gamma^\vee\cdot q_\beta\,=\,0~, 
\label{gg}
\end{align}
where $\alpha,\beta,\gamma\in\Delta$ are simple roots, $\alpha$ is a long root, while $\beta,\gamma$ are short roots. An obvious solution to  (\ref{bb}),(\ref{gg}) is 
$q_\beta=-m \beta^\vee$, which complies with   (\ref{aid}). To see that this is the only possible solution introduce the electric charges $x_\beta$ via $q_\beta=-m\beta^\vee+x_\beta$. Then   (\ref{bb}) gives
\begin{equation}
	\alpha\cdot x_\beta=0~.
	\label{xx}
\end{equation} 
Fixing $\beta$ here and running $\alpha$ over the set of all long simple roots one immediately concludes that $x_\beta=0$, which leads to the desired identity $\CG_\beta=\CG_{\beta,\,m}$. Thus we verified that if there is one massless dyon with the charge $\CG_{\alpha,\,m}$, then there exist a set of $r$ dyons with charges satisfying   (\ref{aid}).

Combining  (\ref{Gam}),(\ref{aid}) one concludes that in order to describe the breaking of the supersymmetry $\CN=2\rightarrow\CN=1$ in terms of light dyons it is necessary to consider a particular set of dyons, call it the  minimal set, which have the following charges
\begin{equation}
	\CG_{\,\alpha, \,m}=(\,\av,-m\,\av\,)~, \quad \quad\quad \alpha \, \in \,\,	\Delta~,\quad m\,\in\,Z_{\,\hv}~.
	\label{ma}
\end{equation}
For a given $m$ this set includes a subset of $r$ dyons. Each one of them becomes massless at some value of the scalar field. Generically these values are different for different dyons. However, there should exist a particular value of the scalar field that makes all these $r$ dyons massless simultaneously. A necessity for this phenomenon follows from the fact that the vacuum of the $\CN=1$ supersymmetric gauge theory is obviously invariant under the Weyl reflection. Consequently, the condensate of dyons that describes a transition $\CN=2\rightarrow \CN=1$ should be invariant under the Weyl  reflections as well. This condition can only be satisfied if all $r$ dyons participate in the condensate becoming massless simultaneously.

Different values of $m$ in   (\ref{ma}) correspond to different vacua of the $\CN=1$ theory. 
The number of different vacua of the $\CN=1$ supersymmetric gauge theory equals its Witten index specified in    (\ref{I=h}), $I_\mathrm{\,W}=\hv$.  A shift of $m$ in the set (\ref{ma}) is generated by the chiral transformation of the dyon charge, which is described by  (\ref{G->GM}),(\ref{gqgq}). This transition is matched by the chiral transformation of the gaugino condensate in   (\ref{ll}). 

Remember that generically   (\ref{qdirect}) states that electric charges are allowed to reside anywhere in the lattice of roots. Equation (\ref{ma}) is more assertive, stating that for the dyons considered the electric charges are collinear to their magnetic charges and lie in the lattice of coroots, which for nonsimply-laced gauge algebras incorporates fewer points than the lattice of roots. 

The minimal set of dyons (\ref{ma}) allows different representations. The definition of the set of simple roots $\Delta$  depends on the choice of the basis in the $r$-dimensional space of roots. Taking a different basis, one gets a different minimal set of light dyons. \footnote{This implies that if $\alpha$ is any root, $\alpha\in \mathsf{g}$, then the dyon with the charge $\CG_{\,\alpha,\,m}$ is also light.} An expansion of this construction can be made using transformations of charges of light dyons given by a symplectic, integer valued  matrix
\begin{equation}
	\CG\, \rightarrow\, \CG^{\,'}\,=\,\CG\,M~,\quad M\,\in\, Sp\,(2r,Z)~,
	\label{MIII}
\end{equation}
which is in line with the idea known as the democracy of dyons. Considering this transformation, one needs to transform simultaneously the subset of those $r$ dyons, which are massless at some particular value of the scalar field. 
We will use this option in Section \ref{comp} to verify that a set of light dyons, which at first sight looks different from the minimal set, can be made compliant with   (\ref{ma}).

Let us briefly repeat the arguments, which allow one to unfold the minimal set of light dyons
specified in   (\ref{ma}) starting from one monopole. Take a light monopole with the charge $\CG_{\,\alpha, \,0}$. Using the Weyl reflections verify that  there should exist also light monopoles with the charges $\CG_{\,\beta,\,0}$, $\beta\in\Delta$, overall $r$ of them. Applying the chiral transformations to the charges $\CG_{\,\beta,\,0}$ reproduce the full set of $\hv\, r$ charges of light dyons in   (\ref{ma}). One learns that $\hv\, r$ is the minimal number of light dyons necessary to describe the condensates of dyons responsible for the breaking of the supersymmetry $\CN=2\rightarrow\CN=1$. 

These arguments make it tempting to presume that the minimal set of light dyons specified by   (\ref{ma}) incorporates all light dyons. For future reference let us call this assumption the minimal hypothesis, or conjecture. Let us reiterate the facts that support its validity. The presented derivation of the minimal set of dyons uses only basic, fundamental symmetries of the theory, the discrete chiral symmetry and the Weyl symmetry. The minimal set as able to reproduce the transition from the $\CN=2$ supersymmetric gauge theory to the $\CN=1$ case, describing important features of the $\CN=1$ supersymmetric gauge theory, the mass gap and the number of vacua. 
%The following discussion provides several tests of the conjecture.

\section{Strong and weak coupling monodromies}
\label{braid}
It was argued in Refs. \cite{Klemm:1994qj,Argyres:1995jj,Klemm:1995wp} that 
every light dyon generates a monodromy
\begin{equation}
M(\CG)=1+\Omega\,\CG^{\,T}\otimes\,\CG\,=\,
\begin{pmatrix} ~~1+q\otimes g & \quad\, q\otimes q  \\  \quad~\! -g\otimes g & 1-g\otimes q~~\end{pmatrix}~,
	\label{M(G)}
\end{equation}
where $\CG=(g,q)$ is the dyon charge.  This monodromy satisfies the duality condition
\begin{equation}
	M(\CG)\,\Omega\, M(\CG)^T\,=\,\Omega~.
	\label{dua}
\end{equation}
Combined with   (\ref{M(G)}) it implies $M(\CG)\in Sp\,(2r,Z)$. The charge ${\mathcal G}$ of the dyon is an eigenvector of this monodromy with the eigenvalue 1 
\begin{equation}
	\CG \,M(\CG)\,=\,\CG~.
	\label{eigen1}
\end{equation}
To illustrate  validity of   (\ref{M(G)}) remember the gauge group $\mathsf{G}=SU(2)$ discussed in  \cite{Seiberg:1994rs}, which has two light dyons with the charges defined in   (\ref{SU2}). According to (\ref{M(G)}) these charges generate the following monodromies
\begin{eqnarray}
M(\,\CG_0\,)	=
\begin{pmatrix} ~~\,1 & 0~  \\  -2 & 1~\end{pmatrix}~,\quad\quad
M(\,\CG_{1}\,)	=\begin{pmatrix} \,-1 & 2~  \\  \,-2 & 3~\end{pmatrix}~.
	\label{M01}
\end{eqnarray}
For $\mathsf{G}=SU(2)$ there is only one simple root $\alpha$, and consequently only one matrix ${R}\equiv {R}_\alpha$, which represents a monodromy at weak coupling   (\ref{Ra})
\begin{equation}
	{R}\,=\,
	\begin{pmatrix} -1 & ~~2\,  \\ ~~\,0 & -1 \,\end{pmatrix}~.
	\label{Ra2}
\end{equation} 
Equations (\ref{M01}),(\ref{Ra2}) imply an important equality
\begin{equation}
	M(\,\CG_0\,)\,M(\,\CG_{1}\,)\,=\,{R}~,
	\label{MMR}
\end{equation}
that was casted as $M_1M_{-1}=M_\infty$ in \cite{Seiberg:1994rs}, which matches the monodromies related to light dyons at strong coupling with the monodromy at weak coupling.

Consider a generalization of   (\ref{MMR}) for an arbitrary gauge group.  The  monodromy (\ref{M(G)}) for a light dyon with the charge $\CG_{\,\alpha \,m}$ reads
\begin{equation}
	M\left(\,\CG_{\,\alpha, \,m}\,\right)\,\equiv\,
	M_{\,\alpha,\,m}\,=\,
	\begin{pmatrix} ~1-m \,\av \otimes \,\av &\quad m^2\, \av\otimes \,\av  
	\\ \quad\	~\,\, -\av \otimes \,\av & 1+m\;\av\otimes \,\av~\end{pmatrix}~.
	\label{finds}
\end{equation}
If $\alpha$ is a long root, $\alpha^2=2$, then   (\ref{finds}) implies that
\begin{equation}
M_{\,\alpha,\,m}\,M_{\,\alpha,\,m+1}\,=\,
{R}_{\,\alpha}~,
	\label{lon}
\end{equation}
which rephrases   (\ref{MMR}). This means that for any simply laced $\mathsf{g}$   (\ref{lon}) gives the desired relation between the monodromies at strong and weak couplings.

For nonsimply laced gauge groups we consider below two possibilities. The first one presumes that the minimal hypothesis, which was formulated at the end of Section \ref{minimal}, is valid. We will see that in this case   (\ref{lon}) needs to be modified for short roots.  The other option is to discard the minimal hypothesis extending the minimal set of light dyons. Then the condition that matches the strong and weak coupling monodromies can be presented in a form similar to   (\ref{lon}). 

\subsubsection{Minimal hypothesis and basic monodromies for light dyons}
\label{min-basic}
Consider the first opportunity, assuming that the minimal hypothesis is valid, i.e. the minimal set specified in   (\ref{ma}) exhausts the list of light dyons. In this case it is convenient to introduce a new matrix $\mathcal{M}_{\,\alpha,\,m}\,\in\,Sp\,(\,2r,Z\,)$ 
\begin{equation}
\mathcal{M}_{\,\alpha,\,m}=\bfone+\frac{1}{\nu_\alpha}(M_{\,\alpha,\,m}-\bfone)
	=
		\begin{pmatrix} 1-m \,\alpha \otimes \,\av & \quad ~m^2\, \alpha\otimes \,\av  
		\\  
		\quad~\,-\alpha \otimes \,\av & 1+ m\,~\alpha\otimes \,\av~\end{pmatrix}\,.
	\label{Mnu}
\end{equation}
Here $\bfone$ is the $2r\times 2r$ unity matrix, and
\begin{equation}
	\nu_\alpha\,=\,2/\alpha^2\,=\,1,2,3~,
	\label{123}
\end{equation}
which gives $\nu_\alpha=1$, for long roots, $\nu_\alpha=2$ for short roots of any algebra except $G_2$, and $\nu_\alpha=3$ for short roots of $G_2$.  Equation (\ref{finds}) implies that $({M}_{\,\alpha,\,m}-\bfone)^2=0$, which leads to
\begin{equation}
\big(\,\mathcal{M}_{\,\alpha,\,m}\,\big)^{\nu_\alpha}\,=\,{M}_{\,\alpha,\,m}~.
	\label{M^nu}
\end{equation}
The matrix $\mathcal{M}_{\,\alpha,\,m}$ possesses the following properties 
\begin{align}
	\mathcal{M}_{\,\alpha,\,m}\,\Omega\, \mathcal{M}_{\,\alpha,\,m}^T\,=\,\Omega~,
	\label{duaNew}\\
	\CG_{\,\alpha,\, m}\,\mathcal{M}_{\,\alpha,\,m}\,=\,\CG_{\,\alpha, \,m}~,
	\label{GM=G}
\end{align}
which are similar to  (\ref{dua}),(\ref{eigen1}) that define the conventional monodromy related to a light dyon.  Equations (\ref{duaNew}),(\ref{Mnu}) imply $\mathcal{M}_{\,\alpha,\,m}\in
Sp\,(2r,Z)$. Let us call $\mathcal{M}_{\,\alpha,\,m}$ the basic monodromy related to the light dyon; it is basic in the sense that the conventional monodromy ${M}_{\,\alpha,\,m}$ equals its integer power, as per   (\ref{M^nu}). In particular, for a long root $\alpha$ they coincide, $\mathcal{M}_{\,\alpha,\,m}={M}_{\,\alpha,\,m}$.

From   (\ref{Mnu}) one finds that $\mathcal{M}_{\,\alpha,\,m}$ satisfies the following identity
\begin{align}
	\mathcal{M}_{\,\alpha,\,m}\,\,\mathcal{M}_{\,\alpha,\,m+1}\,=\,
\mathcal{R}_{\,\alpha}~.	
	\label{MnuMnuR}
\end{align}
Here the matrix
\begin{equation}
\mathcal{R}_{\,\alpha}\,=\,\begin{pmatrix}
~\rho_\alpha & ~\alpha \,\otimes\,\av ~\\ ~0 & \rho_\alpha
\end{pmatrix}	\,=\,P_\alpha\,\big(\,T_\alpha\,\big)^{\nu_\alpha}\,=\,R_{\,\alpha}\,\big(\,T_{\,\alpha}\,\big)^{\,\nu_\alpha-1}~,
	\label{NewR}
\end{equation}
where $P_\alpha$ and $T_\alpha$ are defined in  (\ref{Rrho}),(\ref{Ta}). The matrix 
$\mathcal{R}_{\,\alpha}$ has a similarity with the matrix of the monodromy $R_{\,\alpha}$ defined in   (\ref{Ra}). The difference between them is in an integer coefficient  in front of the nondiagonal matrix element in the $2\times 2$ block matrices, which equals $\alpha\otimes\alpha$ in (\ref{Ra}), and $\alpha\otimes\av=\nu_{\alpha}\, \alpha\otimes\alpha$ in (\ref{NewR}). For a long root $\alpha$  this difference disappears, 
$\mathcal{R}_\alpha = {R}_\alpha $

Consider a short root $\alpha$ for a nonsimply laced $\mathsf{g}$, when $\mathcal{R}_\alpha \ne {R}_\alpha $. Remember that discussing the monodromy $R_{\,\alpha}$ for the scalar field $\Phi$ we used a path that connects $A$ with $\rho_\alpha A$. Equation (\ref{Ra}) was written presuming the simplest path, which crosses only once the wall of the Weyl camera that is orthogonal to $\alpha$. This path gives the contribution $-\pi i$ to the logarithmic function in   (\ref{APTh}), which leads to the nondiagonal term $ \alpha \otimes \alpha$ in the block-matrix defining $R_{\,\alpha}$ in (\ref{Ra}). One can modify the path that connects $A$ with $\rho_\alpha A$ in such a way as to reproduce $\mathcal{R}_\alpha $ instead. Assume that $\alpha$ is a short root of $\mathsf{G}$, $\mathsf{G}\neq G_2$, when $\nu_\alpha=2$. Consider the path that starts from $A$ and crosses the same wall of the Weyl camera twice,
forward and backward. Presume that this double crossing results in the combined variation of the logarithmic function $-2\pi i$. Assume the path returns back to the starting point $A$. After that take the Weyl reflection $P_\alpha$, which brings the path to the desired final point $\rho_\alpha A$. This definition of the path leads to the monodromy $\mathcal{R}_{\,\alpha}$ for the scalar field $\Phi$. 
Similarly, when $\mathsf{G}=G_2$ and $\nu_\alpha=3$, consider the path which runs from $A$ to $\rho_\alpha\,A$ crossing the wall of the camera trice, twice forward and once backward, giving the total contribution $-3\pi i$ to the logarithmic function. This path again leads to the monodromy $\mathcal{R}_{\,\alpha}$ in   (\ref{NewR}). We see that $\mathcal{R}_{\,\alpha}$ and $R_{\,\alpha}$ describe the same property of the system, the monodromy at weak coupling, which is defined by the path along which the variation of the scalar field is taken. The difference between them is related to the way this path is defined for short roots.

We conclude that embracing the minimal hypothesis, one can rely on   (\ref{MnuMnuR}), which generalizes   (\ref{MMR}) and provides a match between the monodromies at strong and weak couplings. The existence of this match supports the validity of the minimal hypothesis. However, to make this support stronger, one needs to verify that the basic monodromies are compatible with the Seiberg-Witten solution.

\subsubsection{Including more light dyons}
\label{extension}
Consider the alternative option. Let us extend the set of dyons presuming that alongside the minimal set (\ref{ma}) there exist also additional light dyons with the charges 
\begin{equation}
	\CG{\,'}_{\!\!\alpha,\,m}=(\,\av,\,-m\,\av\!-\alpha\,)~.
	\label{sho}
\end{equation}
For a long root $\alpha$, $\av=\alpha$, this assumption does not change the set of light dyons because $\CG{\,'}_{\!\!\alpha,\,m}=\,\CG_{\,\alpha,\,m+1}$. Therefore for the simply laced $\mathsf{g}$   (\ref{sho}) does not define new charges. However, for short roots $\alpha$ (\ref{sho}) makes a presumption that there exist light dyons with new charges, thus enlarging the set of light dyons. From   (\ref{M(G)}) one derives
\begin{equation}
		M\left(\CG{\,'}_{\!\!\alpha, \,m}\right)
		\equiv
		  M^{\,'}_{\alpha, \,m}=
	\begin{pmatrix} 1-(\,m\av+\alpha\,)\, \otimes \,\av &  (\,m\av+\alpha\,)\otimes \,(\,m\av+\alpha\,)~  
	\\ \quad\quad\quad\quad\quad -\av \otimes \,\av & \quad~~~1+\av\,\otimes \,(\,m\av+\alpha\,)~\end{pmatrix},
\label{M'}
\end{equation}
and consequently finds
\begin{equation}
	M_{\,\alpha,\,m}\,M{\,'}_{\!\!\alpha,\,m}\,=\,{R}_{\,\alpha}~.
\label{MM'}
\end{equation}
This relation generalizes   (\ref{MMR}) for an arbitrary gauge group by using only conventional monodromies $M(\CG)$, though this construction presumes that the minimal set of dyons is extended.

For the gauge group $G_2$,  there exists an additional ambiguity. Alongside the dyons with charges given by   (\ref{sho}) one can  presume also/instead that there exist light dyons with charges 	
\begin{equation}
\CG{\,''}_{\!\!\alpha,\,m}=(\,\av,\,-m\,\av+\alpha\,)~, 	
	\label{G''}
\end{equation}
which differ from $\CG{\,'}_{\alpha,\,n}$ for any values of $m,n$, $\CG{\,''}_{\!\!\alpha,\,m}\neq
\CG{\,'}_{\!\alpha,\,m}$ due to the mere fact that short roots of $G_{\,2}$ satisfy $\av=3\alpha$. 
\footnote
{This ambiguity does not manifest itself for other nonsimply laced gauge algebras, since their small roots satisfy $\av=2\alpha$ leading to $\CG{\,''}_{\!\!\alpha,\,m+1}=\CG{\,'}_{\!\!\alpha,\,m}$.}
Introducing 
\begin{equation}
M\left(\CG{\,''}_{\!\!\alpha, \,m}\right)\,=\,M{\,''}_{\!\!\alpha,\,m}~, 	
	\label{M''}
\end{equation}
one can write
\begin{equation}
M{\,''}_{\!\!\alpha,\,m}\,M_{\,\alpha,\,m}\,=\,{R}_{\,\alpha}~, 	
	\label{M''M}
\end{equation}
which gives another possible generalization of   (\ref{MMR}). 

Summarizing, we discussed in Subsections \ref{min-basic}, \ref{extension} two options, which allow one to write a relation between the monodromies at strong and weak couplings. One of them presumes that the minimal set exhausts the list of light dyons, which leads to   (\ref{MnuMnuR}) written in terms of the basic monodromies (\ref{Mnu}). Alternatively, one can rely on conventional monodromies for light dyons in   (\ref{MM'}), but then the list of light dyons should go beyond the minimal set. The difference between these two opportunities manifests itself only for nonsimply laced gauge algebras. To establish which one of the two available options takes place, a more detailed study is necessary.

\section{Comparison with known results}
\label{comp}
In order to test   (\ref{ma}) let us discuss several facts, which have been known for dyon charges previously.
 
\subsection{Gauge group $SU(2)$}
For $\mathsf{G}=SU(2)$, which was discussed in \cite{Seiberg:1994rs},   (\ref{SU2}) shows that the number of light dyons as well as their charges comply with   (\ref{ma}).
	
\subsection{Gauge group $SU(3)$}
For $\mathsf{G}=SU(3)$   \cite{Klemm:1995wp} found the following charges of light dyons 
\footnote{Equation (4.20) of  \cite{Klemm:1995wp} gives the magnetic and electric charges in the simple root basis and Dynkin basis respectively.}
\begin{subequations}
\label{G16}
\begin{eqnarray}
\CG_1 &\,=\,&(\alpha_1,\,-\alpha_1)\,,
\label{G1}
\\
\CG_2 &\,=\,&(\alpha_1,\,0)\,,
\label{G2}
\\
\CG_3 &\,=\,&(\alpha_2,\,0)\,,
\label{G3}
\\
\CG_4 &\,=\,&(\alpha_2,\,\alpha_2)\,,
\label{G4}
\\
\CG_5 &\,=\,&(-\alpha_1-\alpha_2,\,\alpha_1)\,,
\label{G5}
\\
\CG_6 &\,=\,&(-\alpha_1-\alpha_2,-\alpha_2)\,.
\label{G6}
\end{eqnarray}
\end{subequations}
The roots $\alpha_i,~i=1,2$ are labeled conventionally (same is true for the roots of other groups discussed below). To clarify notation let us present them in the Cartesian coordinates $(x,y)$ in the following form
\begin{equation}
\alpha_1\,=\,\frac{1}{\sqrt{2}} \,\,(\,1,-\sqrt{3}\,)~,
\quad 
\alpha_2\,=\,\frac{1}{\sqrt{2}} \,\,(\,1,\,\sqrt{3})~.
\label{alpha}
\end{equation}
Consider the Weyl reflection of the charge $\CG_1$ in the plain that is orthogonal to the root ${\alpha_1+\alpha_2}$, which is accompanied by a change of sign of the charge of the dyon (by taking the anti-dyon). The result reads
\begin{equation}
\CG_1=(\alpha_1,\,-\alpha_1) \,\rightarrow \,-\,\CG_1\,P_{\alpha_1+\,\alpha_2}\,=\,
(\alpha_2,\,-\alpha_2)\,\equiv \,\CG^{\,'}~.
	\label{G1->}
\end{equation}
Make a similar transformation for the charge $\CG_4$
\begin{equation}
\CG_4=(\alpha_2,\,\alpha_2) \,\rightarrow \,-\,\CG_4\,P_{\alpha_1+\,\alpha_2}\,=\,
(\alpha_1,\,\alpha_1)\,\equiv \,\CG^{\,''}~.
	\label{G2->}
\end{equation}
Observe that dyons with charges 
\begin{subequations}
\label{123456final}
\begin{align}
& \{\,\CG_1,\,\CG_2,\,\CG^{\,''}\}\,=\,\CG_{\,\alpha_1,\,m}~,
\label{126}
\\
&\{\,\CG^{\,'},\,\CG_3,\,\CG_4\,\}\,=\,\CG_{\,\alpha_2,\,m}~,
\label{634}
\end{align}
\end{subequations}
comply with the formula for $\CG_{\,\alpha,\,m}$ in   (\ref{ma}) in which $\alpha=\alpha_1,\alpha_2$ and $m\in Z_{3}$. 

At a glance the charges $\CG_5$ and $\CG_6$ in  (\ref{G5}),(\ref{G6}) deviate from predictions of   (\ref{ma}), but there is a way to transform these charges  making them identical to $\CG^{\,''}$ and $\CG^{\,'}$ respectively. In \cite{Klemm:1995wp} it is stated that pairs of mutually local dyons can become massless simultaneously, which agrees with the discussion in Section \ref{minimal}. Consequently the dyons with the charges $\CG_1$ and $\CG_6$ become massless simultaneously. Consider the $6\times 6$ matrix $M$
\begin{equation}
	M=
	\,\begin{pmatrix}
~1 & 0 \\
-g\otimes g  &1
\end{pmatrix} ~.	
	\label{M1}
\end{equation}
where $g$ is a vector 
\begin{equation}
	g\,=\,\frac{1}{\sqrt{3}}\,\,\alpha_1 \,+ \frac{2}{\sqrt{3}}\,\,\alpha_2
	\,=\,\frac{1}{\sqrt{2}} \,(\,\sqrt{3}\, ,\,1\,)~.
	\label{g1}
\end{equation}
Clearly, $M\in Sp\,(6,Z)$. This fact allows us to apply   (\ref{MIII}) to the charges $\CG_1$ and $\CG_6$, transforming them as follows
\begin{align}
& \CG_1\,\rightarrow \CG_1\,M\,=\,\CG_1~, \\
& \CG_6\,\rightarrow \CG_6\,M\,=\,\CG^{\,'}~.
\label{16}
\end{align}
These equations  show that instead of a pair of mutually local light dyons with the charges $\CG_1,\,\CG_6$ one can consider light dyons with the charges $\CG_1,\,\CG^{\,''}$. 
Similarly, instead of the mutually local light dyons with charges $\CG_4,\,\CG_5$ one can consider the light dyons with charges $\CG_4,\CG^{\,'}$. The necessary transformation is
\begin{align}
& \CG_4\,\rightarrow \CG_4\,\tilde{M}\,=\,\CG_4~, \\
& \CG_5\,\rightarrow \CG_5\,\tilde{M}\,=\,\CG^{\,''}~.
\label{45}
\end{align}
Here 
\begin{align}
&	\tilde{M}=
	\,\begin{pmatrix}
~1 & 0 \\
\tilde{g}\otimes \tilde{g}  &1
\end{pmatrix}\,\in\,Sp\,(6,Z) ~,	
	\label{M2}
\\
&	\tilde g\,=\,\frac{2}{\sqrt{3}}\,\,\alpha_1 \,+ \frac{1}{\sqrt{3}}\,\,\alpha_2
	\,=\,\frac{1}{\sqrt{2}} \,\,(\,\sqrt{3}\,,\,-1\,)~.
	\label{g2}
\end{align}
The net effect of  (\ref{45}),(\ref{16}) is a transformation
of the charges $\CG_5, \CG_6$ into $\CG^{\,''},\,\CG^{\,'}$. As a result the set of light dyons (\ref{G16}) is transformed into the set (\ref{123456final}), whose charges comply with   (\ref{ma}). The total number of these dyons, six,  also agrees with (\ref{ma}). The fact that pairs of mutually local dyons can become massless simultaneously is once again in line with the discussion in Section \ref{minimal}.

%This is interesting since this agreement is not obvious at first sight.

\subsection{Dyon charges and integrable systems}
The charges of dyons were discussed in \cite{Hollowood:1997pp} using the results of \cite{Martinec:1995by}, which related the Seiberg-Witten solution to the spectral curve of a periodic Toda lattice. It was found in \cite{Hollowood:1997pp} that there exist two series of dyons with charges $\CG^{(1)}_{\,\alpha}$ and $\CG^{(2)}_{\,\alpha}$, $\alpha\in \Delta$, 
\begin{subequations}
\label{G1G2}
\begin{align}
	\CG^{(1)}_{\,\alpha}&\,=\,(\,\av,\,p_\alpha \,\alpha\,)~,
	\label{G1in}
	\\
	\CG^{(2)}_{\,\alpha}&\,=\,(\,\av,\,(p_\alpha+1) \,\alpha\,)~.
	\label{G2in}
\end{align}
\end{subequations}
Here $p_\alpha$ are integers, which should be found from equations formulated in 
\cite{Hollowood:1997pp}. These equations are involving, and their general solution has not been presented. There is a similarity between  (\ref{G1G2}) and   (\ref{ma}), in that in both cases the electric and magnetic charges are collinear. There are also distinctions. Equation (\ref{ma}) specifies that the only restriction on the integer $m$ is $m\in Z_{\,\hv}$, and that $m$ is a factor in front of the coroot $\av$, whereas $p_\alpha$, which plays a similar role,  is a factor in front of the root $\alpha$. For more detail see Subsections \ref{su4}, \ref{g22}.

\subsection{Gauge group $SU(4)$}
\label{su4}
For $\mathsf{G}=SU(4)$ it was found in \cite{Hollowood:1997pp} that
\begin{subequations}
\label{G12123}
\begin{align}
\CG^{(1)}_{\alpha_1}&=(\alpha_1,m\,\alpha_1),~\,
\CG^{(1)}_{\alpha_2}=(\alpha_2,(m+1)\alpha_2),~\,
\CG^{(1)}_{\alpha_3}=(\alpha_3,(m+2)\alpha_3),\\
\CG^{(2)}_{\alpha}&=\CG^{(1)}_{\alpha}+(\,0,\,\alpha),\quad\quad 
\alpha= \alpha_1,\,\alpha_2,\,\alpha_3.
\end{align}
\end{subequations}
These charges look similar to the ones in   (\ref{ma}). There is a subtlety though. Equations (\ref{G12123}) presume that $m\,\in\,Z$, while (\ref{ma}) states that $m\,\in\,Z_{\,\hv}=Z_{4}$; $\hv=4$ for $\mathsf{G}=SU(4)$. A possible explanation for this distinction is that the present work aims at finding the smallest possible number of dyons that are necessary to describe the $\CN=2$ gauge theory, while 
  \cite{Hollowood:1997pp} did not pursue this goal.

\subsection{Gauge group $G_2$}
\label{g22}
For $\mathsf{G}\,=\,G_2$  \cite{Hollowood:1997pp} found that there are dyons with charges
\begin{subequations}
\label{G212}
\begin{align}
\CG^{(1)}_{\alpha_1}&=(\av_1,m \alpha_1),\quad\quad~~~ \CG^{(1)}_{\alpha_2}=(\av_2,(3m\!-\!1)\alpha_2)=(\av_2,m\av_2\!-\!\alpha_2),
\\
\CG^{(2)}_{\alpha_1}&=(\av_1,(m+1) \alpha_1),~\, \CG^{(2)}_{\alpha_2}=(\av_2,3m\alpha_2)=(\av_2,m\av_2).
	\label{G12}
\end{align}
\end{subequations}
It is taken into account here that $\av_2=3\alpha_2$. Note a similarity of 
$\CG^{(1)}_{\,\alpha_1}$, $\CG^{(2)}_{\,\alpha_1}$ and $\CG^{(2)}_{\,\alpha_2}$
with $\CG_{\,\alpha,\,m}$ in   (\ref{ma}), though again an integer $m$ is understood differently, $m\in Z$ in  (\ref{G212}), while $m\in Z_{\,\hv}=Z_4$ in   (\ref{ma}), remember $\hv=4$ for $\mathsf{G}=G_2$. The charge  $\CG^{(1)}_{\,\alpha_2}$ does not fit into the minimal set given by   (\ref{ma}), but it complies with an extension of the minimal set discussed in   (\ref{sho}). 
\footnote{The ambiguity for $\mathsf{G}=G_2$ mentioned in  (\ref{sho}),(\ref{G''}) of the present work does not seem to be present in equations (6.10) and (5.12) of   \cite{Hollowood:1997pp}, from which  (\ref{G212}) of the present work are derived.}
%, though a possible inconsistency of signs in some equations in the preprint of %\cite{Hollowood:1997pp} does not allow one to draw a definite conclusion.} 
This latter fact seems to indicate that for $\mathsf{G}=G_2$ the minimal hypothesis proves erroneous, though further study of this point is necessary.

\section{Conclusions}
\label{conc}
It is shown that in the $\CN=2$ supersymmetric gauge theory there exists a minimal set of light dyons with charges specified by   (\ref{ma}). This conclusion is derived from symmetries of the $\CN=2$ and $\CN=1$ supersymmetric gauge theories. One of them is the discrete chiral symmetry. Another is the Weyl symmetry, which is a remnant of the broken gauge symmetry. The minimal set includes $\hv\,r$ light dyons. Here $\hv$ is the dual Coxeter number, which counts the number of vacua of the $\CN=1$ supersymmetric gauge theory and also equals the Witten index for this theory, while $r$ is the rank of the gauge group.

Since the minimal set of dyons accounts for the fundamental symmetries of the problem, it is tempting to conjecture that this set provides a complete list of all light dyons, which are necessary to describe the properties of the $\CN=2$ supersymmetric gauge theory. The validity of this conjecture was verified using the condition that matches monodromies at weak and strong couplings. The conjecture passes the test provided the Seiberg-Witten solution incorporates the so called basic monodromies, which are related to conventional strong-coupling monodromies by   (\ref{M^nu}). This fact makes it interesting to check whether the basic monodromies are present in the Seiberg-Witten solution. The conjecture was also tested by comparison with the known charges of dyons for several simplest gauge groups. The unitary groups $SU(n)$ for $n=2,3,4$  were found to comply with the conjecture, while for $G_2$ the validity of the conjecture is not evident, though this latter case needs to be studied further. 

\vspace{1cm}
%\acknoledgment
A financial support of the Australian Research Council is acknowledged.

% \appendix

% \section{}
% \label{}

\end{document}